\documentstyle[12pt,a4]{article}
\begin{document}

\title{ THE THREE-BODY CONTINUUM COULOMB PROBLEM AND THE 3$\alpha$ 
STRUCTURE OF $^{12}$C}
\author{D.V.~Fedorov \\ European Centre for Theoretical
Studies in Nuclear Physics \\ and Related
Areas,  I-38050 Trento, Italy \\ and \\ A.S.~Jensen \\
Institute of Physics and Astronomy, \\
Aarhus University, DK-8000 Aarhus C, Denmark}
%\date{~}

\maketitle

\begin{abstract}
We introduce an approach, based on the coordinate space Faddeev
equations, to solve the quantum mechanical three-body Coulomb problem
in the continuum. We apply the approach to compute measured properties
of the first two $0^+$ levels in $^{12}$C, including the width of the
excited level, in a 3$\alpha$-particle model. We use the two-body
interaction which reproduces the low energy $\alpha$-$\alpha$
scattering data and add a three-body force to account for the compound
state admixture. \\

\end{abstract}

\paragraph*{Introduction.}
The quantum mechanical three-body problem with Coulomb interaction is
yet one of the unsolved problems of few body physics (see,
for example, \cite{friar,kie94,fri95,lin95}). A number of three-body
Coulomb problems in nuclear, atomic and molecular physics cannot be
dealt with at the moment because of the principal difficulties in
constructing correct asymptotic wave functions in the case when none of
the two body subsystems is bound.

One of such problems is the description of the astrophysically
significant second $0^+$ state in $^{12}$C (denoted $0^+_2$) as a
system of three alpha particles \cite{lan94,ober94}. Despite the long
history of investigations it is yet unclear to what extent the
three-body picture accounts for the real structure of this state as it
lies above the threshold and requires therefore correct continuum
spectrum solutions of the three-body Coulomb problem.

The purpose of this letter is two-fold. First we introduce an approach
to compute the three-body continuum spectrum for Coulomb interactions.
We construct the asymptotic functions employing the adiabatic
hyperspherical expansion of the coordinate space Faddeev equations
\cite{fed93}. This method proved to be very powerful in treating the
long-distance behavior of three-body systems including the pathological
Efimov effect \cite{fed94}. The approach is a generalization of the
adiabatic hyperspherical method, successfully  used in atomic physics
\cite{chr84}, into the more versatile Faddeev equations.

Secondly, we apply the invented method to the two lowest $0^+$ states
in $^{12}$C where descriptions in terms of three interacting
$\alpha$-particles traditionally are believed to be rather accurate.
The second state $0_2^+$ is unbound with respect to three
free $\alpha$-particles and its structure (along with the ground state
structure) has been debated for many years, see for example
\cite{vis71,val76,mar82}. A correct description, which must include
continuum effects of the long-range Coulomb interaction, is therefore
now possible.

As repeatedly suggested we shall include a genuine three-body force in
order to account for three-body effects related to the Pauli principle
and polarization effects \cite{nob70}. Beside providing information
about the structure, the method also enables us to give an estimate of
the width of the $0_2^+$-level within the genuine three-body model.

\paragraph*{Method.}
We solve the Faddeev equations in coordinate space using the
adiabatic hyperspherical expansion method \cite{fed93,fed94b}. We
shall use hyperspherical coordinates, which consist of one radial
coordinate $\rho$ (hyperradius) and five generalized angles denoted by
$\Omega$. The hyperradius $\rho$ is defined in terms of the center of
mass coordinates of the $\alpha$-particles $r_i$ as $ \rho^2 = A_\alpha
\sum_{i=1}^{3} r_i^2$, where $A_\alpha = m_\alpha / m = 3.97$ is given
by the nucleon mass $m$ and the $\alpha$-particle mass $m_\alpha$.

For a fixed $\rho$ we consider the eigenvalue equation for the
five dimensional spin-angular part of the Faddeev operator, i.e.\
\begin{equation}\label{fadeq}
\Big(T_\Omega-\lambda_n\Big)\Phi^{JM}_{n,(i)}
+\frac{2m}{\hbar^2}\rho^2 V_i\Phi^{JM}_n=0 \;, i=1,2,3 \;, \;
\Phi^{JM}_n = \Phi^{JM}_{n,(1)}+\Phi^{JM}_{n,(2)}+\Phi^{JM}_{n,(3)} \;,
\end{equation}
where $T_\Omega$ is the angular part of the kinetic energy operator
and $V_i$ is the potential between particles $j$ and $k$, where
$\{i,j,k\}$ is a cyclic permutation of $\{1,2,3\}$. The eigenvalues
$\lambda_n(\rho)$ and the eigenfunctions $\Phi^{JM}_n(\rho,\Omega)$,
which also carry the dependence of the spins of the particles, are
calculated as function of $\rho$. The total three-body wave function
is then expanded in terms of these eigenfunctions, i.e.\
\begin{equation}
\Psi^{JM}=\frac{1}{\rho^{5/2}}\sum_n f_n(\rho)\Phi^{JM}_n(\rho,\Omega).
\end{equation}
 
The system of coupled hyperradial equations for the expansion
coefficients $f_n(\rho)$ finally becomes
\begin{eqnarray}\label{eqr}
&\Big(-\frac{d^2}{d\rho^2}-\frac{2mE}{\hbar^2}
	+{15/4\over\rho^2}+{\lambda_n(\rho)\over\rho^2}\Big) f_n(\rho)=
	\sum_{n'}\Big(2P_{nn'}\frac{d}{d\rho}+
	Q_{nn'}\Big) f_{n'}(\rho) \; , &\\
&P_{nn'}(\rho)=\int d\Omega
\Phi_{n}^*{\partial\over\partial\rho}\Phi_{n'},\;\;
Q_{nn'}(\rho)=\int d\Omega
\Phi_{n}^*{\partial^2\over\partial\rho^2}\Phi_{n'} \; . &  \nonumber
\end{eqnarray}
For large $\rho$ the non-diagonal terms $P$ and $Q$ are small
compared to the diagonal terms $\lambda_n(\rho)$ which, divided by
$\rho^2$, serve as an effective adiabatic hyperradial potential.

The large distance tail of the eigenvalues $\lambda_n(\rho)$ is
independent of the properties of the short-range nuclear force. The
derivation of this asymptotic behavior is difficult and beyond the
scope of this letter, but the result can be written as \cite{fedtobe}
\begin{equation}\label{asym}
\lambda_n(\rho\rightarrow\infty) \rightarrow
3\eta\rho+C_n\sqrt{\eta\rho}+...\;,
\end{equation}
where $\eta={2m\over\hbar^2}(2e)^2\sqrt{A_\alpha}$, $2e$ is the charge
of the $\alpha$-particle and $C_n$ are constants. For the lowest term
this constant is $C_1=9/4$. At large distances the adiabatic potential
$\lambda_n/\rho^2$ has therefore a Coulomb term $3\eta/\rho$, with the
next term behaving as $\rho^{-3/2}$. These terms are absent for
uncharged particles.

For a positive energy $E\equiv\hbar^2\kappa^2/(2m)$ the general
solution of equation~(\ref{eqr}) asymptotically is a linear combination
of two linearly independent solutions $\exp({\pm
i(\kappa\rho-{3\eta\over 2\kappa}\log{\kappa\rho}))}$. A resonance with
energy $E_r$ and width $\Gamma$ corresponds to a solution of the system
(\ref{eqr}) with the complex energy $E_0=E_r-i\Gamma/2$ \cite{landau}.
This solution satisfies the boundary condition
\begin{equation}\label{bndr}
f_n(\rho\rightarrow\infty)\rightarrow
\exp(+i(\kappa_0\rho-{3\eta\over 2\kappa_0}\log{\kappa_0\rho})),
\end{equation}
where $\kappa_0=\sqrt{2m(E_r-i\Gamma/2)/\hbar^2}$. This boundary
condition determines that the $S$-matrix has a pole at the complex
energy $E_0$.

\paragraph*{Interactions.}
We use the Ali-Bodmer potential version "a" \cite{ali66} slightly
altered to reproduce the s-wave resonance of $^{8}$Be at 0.093~MeV in
addition to the low energy $\alpha\alpha$ phase shifts. The potential
is given as 
\begin{eqnarray} V=\Big(
125\hat{P}_{l=0}+20\hat{P}_{l=2} \Big)e^{-r^2/1.53^2}-
	30.18 \; e^{-r^2/2.85^2},
\end{eqnarray} 
where lengths and energies are in units of fm and MeV, respectively.
The $\hat{P}_{l}$ is the projection operator onto the state with
relative orbital angular momentum $l$.

It is well known that this potential underbinds the $0^+$ states
\cite{vis71} due to the strong repulsive core which is necessary to
simulate the Pauli repulsion in the 2$\alpha$ system. The width of the
resonance depends crucially on its energy, and to estimate this width
accurately we need the resonance energy at the correct position. We
achieve this by indtroducng an additional attraction.

Indeed, when three $\alpha$-particles are close together one of them
polarizes the other two beyond the polarization included in the
two-body interaction.  Thus the Pauli repulsion between them should be
reduced. This effect can be imitated by a phenomenological three-body
attractive force, which must be non-vanishing only when all three
particles are close together, that is for small $\rho$.  This ``off
shell'' effect must disappear at large $\rho$ where the ``on shell''
properties of the phase fitted interactions provide the correct
asymptotic behavior. As a novel feature we include this additional
interaction by parametrizing the three-body potential as a simple
gaussian, $V_3(\rho)=S_3$ exp$(-\rho^2/b_3^2)$, which is added directly
in the hyperradial equations (\ref{eqr}).

The strength and the range parameters $S_3$ and $b_3$ are now adjusted
to reproduce the position of the $0^+_2$ level at 0.38~MeV and in
addition to get the ground state energy as close as possible to the
observed value of $-7.27$~MeV. This requirement uniquely specifies the
range and the strength of the three-body force. The resulting parameters
are given in table~1.

\paragraph*{Properties of the system.}
The model is now completely defined and we can compute various
measured properties of these two $0^+$ states. The crucial
quantity is the effective potential in the hyperradial equation,
i.e.
$W_n\equiv\hbar^2/(2m)((\lambda_n+{15\over4})/\rho^2-Q_{nn})$
which is shown in fig.~1. The position of the $0^+_2$ level is
indicated by the dashed line. The combined effect of the
three-body centrifugal and Coulomb forces creates a strong
barrier extending approximately from 13 to 60 fm. This barrier
is responsible for the extremely small width of the resonance.

The {\bf root mean square radius} is in the three-body model given by
\begin{equation}
R_{rms}=\sqrt{R_\alpha^2+\frac{1}{3A_\alpha}\langle\rho^2\rangle}
 \;  , \;  \;
 \langle\rho^2\rangle=\sum_n \int_0^\infty f_n^2(\rho)\rho^2 d\rho  \; ,
\end{equation}
where $R_\alpha = 1.47$~fm is the root mean square radius of the
$\alpha$-particle. As seen in table~1 we underestimate the radius
slightly in contrast to a larger overestimate found in other
three-body calculations. This difference is due to the attractive
three-body force which tends to decrease the size of the system.

The {\bf monopole matrix element} is written as
\begin{equation}
M=2\sum_{i=1}^{3} \langle 0^+_{gs}| r_i^2 | 0^+_2 \rangle
	=\frac{2}{A_\alpha}\langle 0^+_{gs}| \rho^2 | 0^+_2 \rangle
 =\frac{2}{A_\alpha}\sum_n \int_0^\infty 
f_n^{(gs)}(\rho) f_n^{(ex)}(\rho) \rho^2 d\rho \; ,
\end{equation}
where $f_n^{(gs)}(\rho)$ and $f_n^{(ex)}(\rho)$ are the radial
wavefunctions for the ground state and the excited state,
respectively.  The numerical result given in table~1 is found with the
wavefunction for the excited state $0^+_2$ calculated as a true bound
state by using an infinite wall at 30 fm.  We overestimate the matrix
element significantly less than other calculations indicating that we obtain
a rather good description of the spatial configurations of the two
$0^+$ levels.

To estimate the {\bf width} of the $0^+_2$ state we first use the
quasiclassical approximation, that is calculate the penetrability of
the combined centrifugal and Coulomb barrier in the lowest adiabatical
channel. The penetration factor ($\sim 10^{-4}$) is multiplied by the
number of assaults per unit time on the barrier ($\sim 1~MeV/\hbar$)
computed as the average radial velocity divided by two times the width
of the potential well. In this approximation the width is about 100~eV.
It is already extremely narrow compared to the resonance energy of
0.38~MeV, although still an order of magnitute larger than the
experimental value. The main source of the underestimate is in the
neglect of the higher adiabatical channels, especially the second one
which has an attractive pocket at $\rho\approx$10~fm, see fig.1.

We now find the complex energy solution of the system (\ref{eqr}) with
the boundary condition (\ref{bndr}).  In order to obtain a converged
result we have to integrate the radial equations numerically up to
$\rho\approx$80~fm. As we can solve numerically the Faddeev
equations (\ref{fadeq}) only until $\rho\approx$40~fm, we have to
extrapolate the eigenvalues accoriding to equation (\ref{asym}).
Because of the low-lying $^8$Be resonance the uncertainty in the third
term of the asymptotic expansion leads to an uncertainty in the width
of about 10~eV.  Within this uncertainty the resulting width is
$\Gamma$=20~eV which is remarkably close to the experimental value.

The width of the $0^+_2$ level in $^{12}$C has also been considered
using a two-body $^8$Be+$\alpha$ approximation within the
$\alpha$-cluster model (see, for example, \cite{mohr}). In this
approximation the $^{12}$C first decays into $^8$Be and an
$\alpha$-particle, then $^8$Be in turn decays into two
$\alpha$-particles.  In our genuine three-body approach the decay
proceeds directly into the three-body continuum and the narrow
resonance $^8$Be comes into play implicitly through the structure of
the angular eigenfunctions and eigenvalues.

The {\bf spatial shapes} of the states are somewhat difficult to
visualize due to the symmetry of the three particles. The wavefunction
depends on six coordinates, which conveniently now can be chosen as the
distance between two of the particles, the distance between their
center of mass and the third particle, the angle between these two
directions and three remaining angles defining the orientation of the
total system.  The latter three are rotational degrees of freedom and
they define a plane containing the three $\alpha$ particles. The
wavefunction of a 0$^+$ state is independent of these rotationsl
coordinates.  It depends upon the first three variables, which define a
triangle with $\alpha$ particles in the corners. To visualize the shape
of the state we sample a sequence of these triangles from the density
distribution.  That is in this sequence the number of triangles of a
certain shape is proportional to the absolute square of the
wavefunction of the set of coordinates corresponding to this triangle.

We then calculate the positions of the alpha particles for each set of
coordinates in the sequence and mark them as points on a plot. The
distribution of these points reflects the spatial structure of the
corresponding state. The frame in which the positions of alpha
particles are plotted is chosen in the following way: i) the center of
mass of the system is positioned in the center of the plot; ii) the
system is oriented so that two of the three particles are in the lower
halfplane and the last particle is in the upper halfplane; iii) one of
the two principal axes of inertia is aligned vertically.  Note that the
condition (ii) introduces certain up-down asymmetry. For symplicity the
origin of the system of coordinates on the plot is put at the lower
left corner.

The result is shown in fig.~2 for the ground state, which clearly
resembles an equal side triangle -- a consequence of the fact that the
ground state contains predominantly the lowest and simplest angular
eigenfunction.  The corresponding contour plot for the excited state is
shown in fig.~3.  Unlike the ground state the excited state has larger
contributions from the higher angular eigenfunctions.  The structure is
therefore a complicated interplay between different configurations with
the tendency to add up to a somewhat elongated triangle.

In addition we calculate the maximum probability configuration. The
latter for the ground state is close to an equal side triangle with the
length of the sides equal to 2.98~fm. For the excited state it is a
somewhat elongated triangle with the length of the sides equal to 3.9,
4.2, 6.1~fm.

\paragraph*{Conclusion.}
We have formulated an approach to deal with the three-body Coulomb
problem in the continuum.  The approach is then applied to three
interacting $\alpha$-particles which gives a description of the lowest
two $0^+$ levels in $^{12}$C. The second of these states is in the
3$\alpha$-particle continuum held together by the Coulomb barrier.  We
construct adiabatic hyperradial potentials and estimate the width of
the excited state by solving the system of hyperradial equations for
the complex energy.  The width, estimated for the first time
consistently within the three-body model, is found to be 20~eV which is
in good agreement with the experimental value proving the validity of
the formulated approach.  The model includes a genuine three-body
force, which imitates off-shell effects or residual three-body
interactions.  Furthermore, the model constitutes an improvement over
previous three-body calculations when we compare with measured values
of the root mean square radius and the monopole transition matrix
element.  The spatial structure of the ground state is confirmed to be
predominantly an equal side triangle while the excited state has a much
broader probability distribution with a tendency to look like a flat
triangle.

{\bf Acknowledgments.} One of us (DVF) acknowledges the support from
INFN, Trento, Italy, where part of this work was done.

\noindent{\Large\bf{Figure Captions}}
\begin{list}{}{\setlength{\leftmargin}{18mm}\setlength{\labelwidth}{16mm}
 \setlength{\labelsep}{2mm}}

\item[Figure 1\hfill]
The effective potentials
$W_n=((\lambda_n+15/4)/\rho^2-Q_{nn})\hbar^2/(2m)$ for the three lowest
adiabatic channels as functions of hyperradius. The dashed line
represents the position of the $0^+_2$ resonance.

\item[Figure 2\hfill]
Contour density plot of the ground state described in detail in the
text. The unit on the axes is fm.

\item[Figure 3\hfill]
The same as in fig.2 for the excited state.

\end{list}

\begin{table}[h]
\centerline{Table 1} 

%\begin{center} 
Parameters and properties of the $0^+$ states.  The results of this
work are given in the first row. The results from the calculations in
ref.~\cite{val76} and ref.~\cite{mar82} are given in the next two rows
and the measured values are shown in the last row.  We show the
parameters (strength $S_3$ and range $b_3$) of the three-body
interactions, which lead to the energies $E(0_i^+)$ of the two $0^+$
states measured relative to the threshold of three free
$\alpha$-particles. Finally we show the root mean square radius
$R_{rms}$ of the ground state, the monopole matrix element $M$ and the
width $\Gamma$ of the excited state. ~\\ 
%\end{center}

\hspace*{1.cm} 
\begin{tabular}{|c|c|c|c|c|c|c|c|}
\hline
 & $S_3$, & $b_3$, & E($0^+_{gs}$), &
 $R_{rms}$, & E($0^+_2$), & $\Gamma(0^+_2)$, & M, \\
 & MeV & fm & MeV & fm & MeV & eV & fm$^2$ \\
\hline
this work & -96.8 & 3.9 & -6.81 & 2.36 & 0.38 & 20 & 6.54 \\
\hline
\cite{val76} & - & - & -6.60 & 2.63 & -0.04 & - & 8.18 \\
\hline
\cite{mar82} & - & - & -14 & 2.68 & -3 & - & 9.35 \\
\hline
 exp. & - & - & -7.25 & 2.47 & 0.38 & 8.3$\pm$1.0 & 5.7 \\
\hline

\end{tabular}
\end{table}

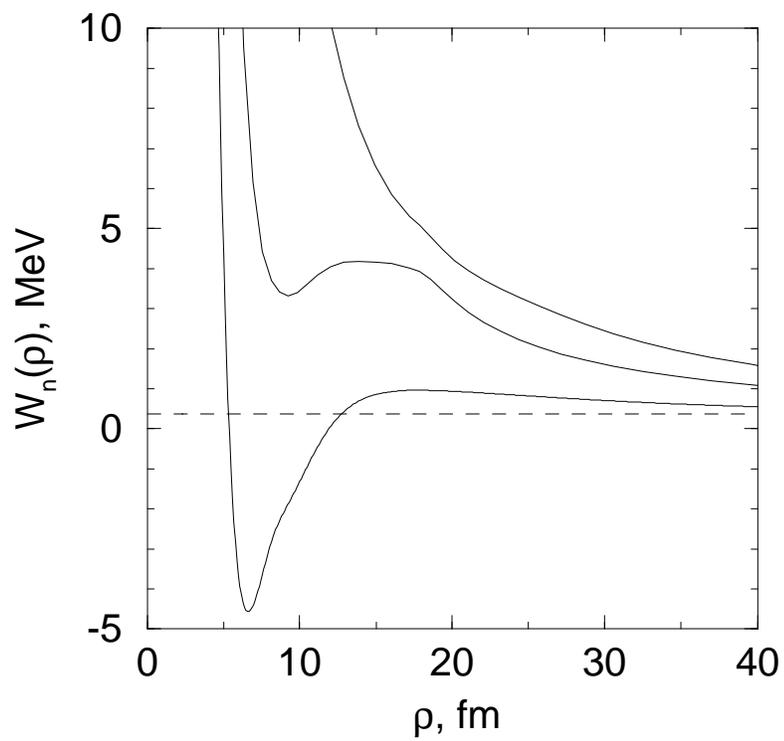
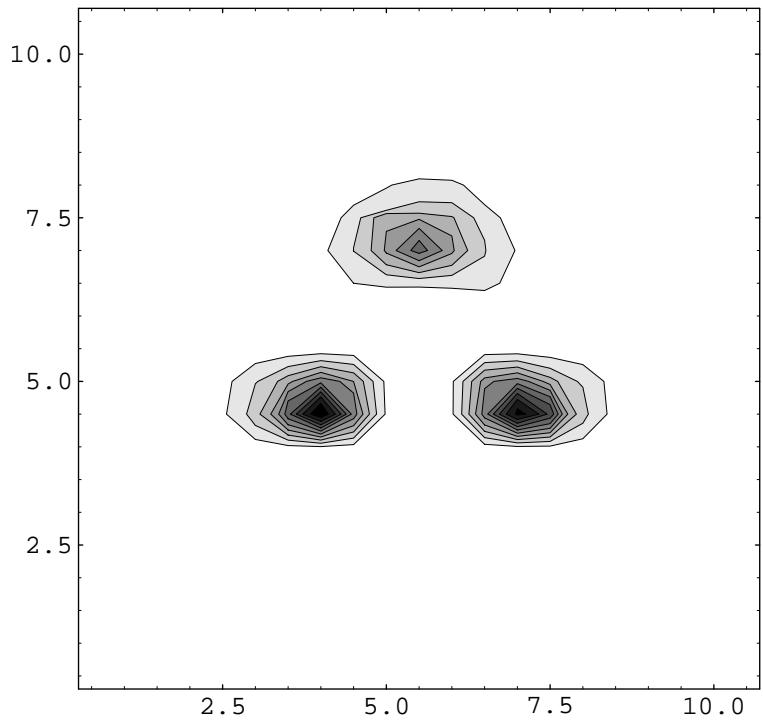
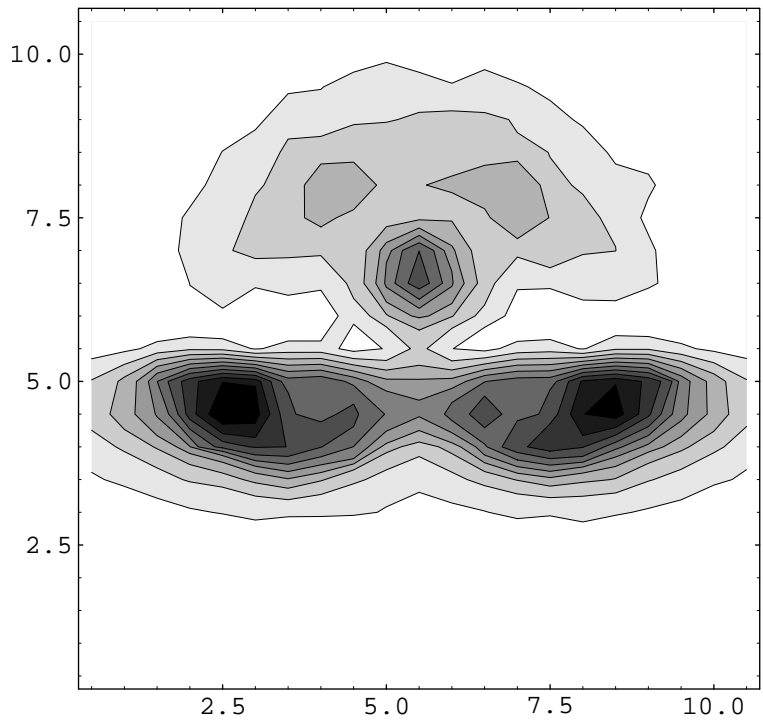
\end{document}